\documentclass[reprint, aps, nofootinbib, prd, draft, floatfix,]{revtex4-2}

\usepackage{amsmath}
\usepackage{amssymb}
\usepackage{tensor}
\usepackage{physics}
\usepackage{xfrac}

\usepackage[autostyle]{csquotes}
\MakeOuterQuote{"}

\usepackage{graphicx} 
\setkeys{Gin}{draft=false} 
\graphicspath{{figures/}}
\usepackage[caption=false]{subfig}

\usepackage[final, hidelinks]{hyperref}
\hypersetup{linktoc=all}
\usepackage{cleveref}

\usepackage{soul}

\newcommand\scri{\mathcal{I}}
\newcommand\half{\frac{1}{2}}
\newcommand\pp{\ensuremath{{\prime\prime}}}
\renewcommand\d{\ensuremath{\partial}}

\newcommand\f{\ensuremath{1 - \frac{2M}{J}}}
\newcommand\fp{\ensuremath{\left(1 - \frac{2M}{J}\right)}}

\newcommand\volelem[1][g]{\sqrt{\abs{#1}}}

\usepackage[obeyDraft, textsize=small]{todonotes}

\begin{document}
\title{Quantum-Corrected Bondi Mass for 2D Hawking Radiation}

\author{Jonathan Barenboim}
\email{jonathan\_barenboim@sfu.ca}
\affiliation{Department of Physics, Simon Fraser University, Burnaby, BC, V5A 1S6, Canada}
\author{Andrei V.\ Frolov}
\email{frolov@sfu.ca}
\affiliation{Department of Physics, Simon Fraser University, Burnaby, BC, V5A 1S6, Canada}
\author{Gabor Kunstatter}
\email{g.kunstatter@uwinnipeg.ca}
\affiliation{Physics Department, University of Winnipeg, Winnipeg, Manitoba, R3B 2E9, Canada}
\affiliation{Department of Physics, Simon Fraser University, Burnaby, BC, V5A 1S6, Canada}

\begin{abstract}
    We derive the Hamiltonian for general semi-classical 2D dilaton gravity, beginning with the complete action including the Polyakov action and Gibbons-Hawking-York boundary term. The value of the Hamiltonian yields a generalized Brown-York quasi-local mass function, and the ADM and Bondi masses are obtained in the appropriate limits. The Bondi mass is equal to the classical mass plus a correction term given by the transformation between initial and final inertial frames. We test the expression for the Bondi mass in the RST model, which can be treated analytically, and in several other models numerically. We find it is monotonically decreasing and remains positive throughout the evaporation process for asymptotically flat black holes.
\end{abstract}

\maketitle
Since the introduction of the CGHS model by Callan et. al. \cite{CallanEtAlEvanescentBlack1992}, two-dimensional dilaton gravity has been a useful toy model for gaining insight into the behavior of quantum effects in gravity and in particular the evaporation of black holes through Hawking radiation. A surprising result commonly found in the literature is that the Bondi mass of an evaporating system can become negative, suggesting that more energy was carried away by Hawking radiation than the black hole contained initially \cite{TadaUeharaConsequencesHawking1995,KimLeeAdmMass1996, KimLeeHawkingRadiation1995,AshtekarEtAlEvaporationTwodimensional2011}. Moreover, in regular black hole models the traditional Bondi mass can be increasing at times, indicating negative outgoing flux at future null infinity. This unusual behavior is a consequence of using a mass defined from the classical theory; that is from the dilaton gravity sector alone, not including the semi-classical extension that incorporates Hawking radiation. 

To remedy the unexpected behavior of the mass, Ashtekar, Taveras, and Varadarajan (ATV) \cite{AshtekarEtAlInformationNot2008} proposed a corrected Bondi mass for the CGHS model obtained from the equations of motion for semi-classical CGHS with a trace anomaly. The ATV Bondi mass is manifestly non-increasing and was shown to remain positive in simulations of evaporating CGHS black holes \cite{AshtekarEtAlEvaporationTwodimensional2011}. 

In this paper we examine the mass for general 2D semi-classical evaporating black holes from a different perspective, deriving the mass through Hamiltonian analysis of the complete semi-classical theory consisting of a generic dilaton gravity action and the Polyakov action~\cite{PolyakovQuantumGeometry1981}. The mass is found to be equal to the classical mass plus a quantum correction term which we show to be equivalent to the ATV mass. We find that in general the corrected Bondi mass is strictly non-increasing and remains positive in numerical simulations of different asymptotically flat black hole models, including both singular and regular black holes, and in the RST model.

Similar approaches have been used to derive a quantum corrected mass in the specific case of the CGHS and RST models, e.g. \cite{BilalKoganHamiltonianApproach1993,deAlwisTwodimensionalQuantum1994,Muller-KirstenEtAl2DQuantum1995}. Early investigations expressed the Polyakov action in a local form with $\eta = 0$ in \cref{eq:z_conformal}, relegating the non-locality to boundary conditions on the constraint equations. This choice does not affect the equations of motion but omits part of the Polyakov field's contribution from the Hamiltonian and Bondi mass. M\"uller-Kirsten et. al. \cite{Muller-KirstenEtAl2DQuantum1995} identified the importance of considering the complete Polyakov action, but a sign error in the timelike component of the boundary term leads to a result inconsistent with ours. The ATV result \cite{AshtekarEtAlInformationNot2008} was derived from the equations of motion, which are unaffected by this gauge choice, hence their mass formula agrees with ours. 

As discussed by de Alwis \cite{deAlwisTwodimensionalQuantum1994}, the unusual global structure of CGHS and RST, featuring two separate future null infinites, can also lead to unexpected behavior of the mass. In this work 2D dilaton gravity is interpreted as the dimensional reduction of a higher-dimensional spherically symmetric spacetime, and the dynamical system has the same global structure as Minkowski space \cite{BarenboimEtAlNoDrama2024}.

The complete action for semi-classical 2D dilaton gravity is
\begin{equation}
    S = S_{DG} + S_M + S_P + S_{GHY}
\end{equation}
where the terms are, respectively, the dilaton gravity action, matter action, Polyakov action, and Gibbons-Hawking-York boundary term.

The 2D dilaton gravity action can be written in a general form \cite{BarenboimEtAlEvaporationRegular2025}
\begin{equation}\label{eq:DG_action}
    S_{DG} = \half \int \volelem \left[ \Phi(r) R + \Phi^\pp(r) (\nabla r)^2 + \Phi^\pp(r) \right] d^2 x,
\end{equation}
where $\Phi$ is an arbitrary function of a scalar dilaton field $r$ and $R$ is the Ricci curvature scalar. We use units ${G=c=1}$. This is the most general action yielding equations of motion containing at most second derivatives of the metric and dilaton, up to a Weyl transformation, and this particular parameterization is chosen because of its connection to spherically symmetric gravity: beginning with the 4D Einstein-Hilbert action
\begin{equation}
    S_{EH} = \frac{1}{16\pi} \int \volelem[^{(4)}g] \, {}^{(4)} R \, d^4 x
\end{equation}
and a spherically symmetric metric
\begin{equation}
    ds^2 = g_{ab}(x^a) dx^a dx^b + r^2(x^a) \, d\Omega^2,
\end{equation}
the curvature can be decomposed as 
\begin{equation}
    ^{(4)}R = R - \frac{4}{r} \nabla^2 r - \frac{2}{r^2} (\nabla r)^2 + \frac{2}{r^2},
\end{equation}
where $R$ and $\nabla$ are the scalar curvature and covariant derivative associated with the 2D metric $g_{ab}$. Integrating out the angular variables and removing a total divergence gives (up to the boundary term)
\begin{equation}
    S_{EH} = \frac{1}{4} \int \volelem \left[ r^2 R + 2 (\nabla r)^2 + 2 \right] d^2 x.
\end{equation}
Hence spherically symmetric general relativity is a special case of the 2D dilaton gravity action \cref{eq:DG_action} with ${\Phi = \half r^2}$. This connection additionally motivates interpreting the dilaton field $r$ as a radial coordinate.

The static vacuum solution of the action \cref{eq:DG_action} is
\begin{equation}\label{eq:static_metric_tr}
    ds^2 = -\fp dt^2 + \fp^{-1} dr^2,
\end{equation}
where $J(r) = \Phi^\prime (r)$ and $M$ is a constant which, as will be shown later, is equal to the ADM mass of the spacetime.

A general metric can be written in conformal gauge as 
\begin{equation}\label{eq:static_metric_tx}
    ds^2 = e^{2\rho} (-dt^2 + dx^2)
\end{equation}
with the static vacuum solution given by 
\begin{equation}\label{eq:r_prime_classical}
    \frac{dr}{dx} = e^{2\rho} = \f.
\end{equation}
The action for $N$ minimally coupled matter fields is 
\begin{equation}
    S_M = -\half \sum^N_{i=1} \volelem (\nabla f_i)^2 \, d^2 x
\end{equation}
and Hawking radiation is incorporated by adding the localized Polyakov action \cite{HaywardEntropyRST1995},
\begin{equation}
    S_P = \frac{\mu}{2} \int \volelem \left( z R - \half (\nabla z)^2 \right) d^2 x,
\end{equation}
where $z$ is an auxiliary scalar field satisfying 
\begin{equation}\label{eq:z_eom}
    -\nabla^2 z = R,
\end{equation} and $\mu = \dfrac{NG\hbar}{24}$ is a constant that determines the strength of the quantum effects. In conformal gauge $z$ may be written
\begin{equation}\label{eq:z_conformal}
    z = 2 \rho + \eta,
\end{equation}
where $\eta$ is a solution of the wave equation $\nabla^2 \eta = 0$ and captures the non-locality of the Polyakov action as boundary conditions on $z$.

Finally, to make the variational principle well-defined a boundary term must also be added to the action,
\begin{equation}
    S_{GHY} = 2 \epsilon \oint_{\partial V} \volelem[h] \chi K d\Sigma,
\end{equation}
where 
\begin{equation}
    \chi = \half (\Phi + \mu z).
\end{equation}
The notation follows closely that in \cite{PoissonRelativistsToolkit2004}: $\partial V$ is the boundary of the manifold, $h$ is the induced metric on the boundary, $\epsilon = \pm 1$ is the norm of the normal vector to the boundary, and $K$ is the trace of the extrinsic curvature on the boundary. We introduce an ADM parameterization, factoring out the conformal factor,
\begin{equation}
    ds^2 = e^{2\rho} \left( -B^2 dt^2 + (dx + A dt)^2 \right),
\end{equation}
and foliate the spacetime with spacelike surfaces $\Sigma_t$ of constant $t$. The boundary can be decomposed into the spacelike boundaries $\Sigma = \Sigma_i \cup \Sigma_f$, and the timelike boundary $\mathcal{B}$ given as the product of boundaries $S_t$ to $\Sigma_t$. The normal vectors to $\Sigma$ and $\mathcal{B}$ are respectively
\begin{equation} \label{eq:normals}
    n^a = \left( \frac{1}{B} e^{-\rho}, -\frac{A}{B} e^{-\rho} \right), \qquad r^a = \left( 0, e^{-\rho} \right),
\end{equation}
with normalization
\begin{equation}\label{eq:normal_norms}
    n^a n_a = -1, \quad r^a r_a = 1, \quad n^a r_a = 0,
\end{equation}
and the induced metrics are 
\begin{equation}
    h = e^{2\rho}, \qquad \gamma = -B^2 e^{2\rho}.
\end{equation}
We first simplify the action by writing the curvature as
\begin{equation}
    R = -2 \nabla_a m^a, \qquad m^a = n^b \nabla_b n^a - n^a \nabla_b n^b.
\end{equation}
Thus 
\begin{equation}
\begin{aligned}
    \int & \volelem  \chi R \, d^2 x = - 2 \int \volelem \chi \nabla_a m^a \, d^2 x \\ 
    &\begin{aligned} =2 \int \volelem \nabla_a \chi \, m^a \, d^2 x & + 2 \int_\Sigma \volelem[h] \chi m^a n_a \, d\Sigma \\ 
     & -2 \int_{\mathcal{B}} \volelem[\gamma] \chi m^a r_a \, d\Sigma. \end{aligned}
\end{aligned}
\end{equation}
Using the relations in \cref{eq:normal_norms}, along with the completeness relation $g^{ab} = -n^a n^b + r^a r^b$, we have 
\begin{equation}
\begin{split}
    m^a n_a = -n^a n_a \nabla_b n^b = K_\Sigma, \\ 
    \quad m^a r_a = \nabla_b r^b = K_\mathcal{B},
\end{split}
\end{equation}
so that 
\begin{equation}
\begin{split}
    \int \volelem \chi R \, d^2 x =& \, 2 \int \volelem m^a \, \nabla_a \chi \, d^2 x \\
    &- 2 \epsilon \oint_{\partial V} \volelem[h] \chi K \, d\Sigma.
\end{split}
\end{equation}

Therefore eliminating the total derivative from the curvature term cancels the GHY boundary term. In the ADM parameterization the total action is 
\begin{widetext}
\begin{equation}
\begin{split}
    S &= \int \volelem \left[ -2 \dot{\chi} b + 2 \chi^\prime A b + 2 \chi^\prime (B^\prime + B \rho^\prime) + \frac{1}{2} \left(J^\prime(r) (\nabla r)^2 + J^\prime(r) - \frac{\mu}{2} (\nabla z)^2 \right) -\half \sum^N_{i=1} (\nabla f_i)^2 \right] d^2 x \\ 
    &= \begin{aligned}[t]
        \int \Biggl[& -2 \hat{\chi} b - 2 B \chi^\pp + 2 B \chi^\prime \rho ^\prime + \frac{J^\prime(r)}{2} \left(-\frac{1}{B} \hat{r}^2 + B {r^\prime}^2 \right) + \frac{B}{2} e^{2\rho} J^\prime(r) - \frac{\mu}{4} \left( -\frac{1}{B} \hat{z}^2 + B {z^\prime}^2 \right) \\
        &-\half \sum^N_{i=1} \left(-\frac{1}{B} \hat{f}_i^2 + B {f_i^\prime}^2 \right) \Biggr] d^2 x + \int_{\mathcal{B}} 2 \chi^\prime B \, d\Sigma
    \end{aligned}
\end{split}
\end{equation}
\end{widetext}
where a dot and prime denote derivatives with respect to $t$ and $x$ respectively, $\hat{\phi} = \dot{\phi} - A \phi^\prime$, and $b = \dfrac{\dot{\rho} - A \rho^\prime - A^\prime}{B}$. In the second line we have integrated by parts to remove derivatives of $B$. The action is now in a form suitable for Hamiltonian analysis. The conjugate momenta are
\begin{gather}
\begin{aligned}
    \Pi_\rho &= -\frac{2}{B} \hat{\chi}, \\
    \Pi_r &= -b \, J(r) - \frac{J^\prime(r)}{B} \hat{r}, \\
    \Pi_z &= -\mu b + \frac{\mu}{2B} \hat{z}, \\
    \Pi_i &= \frac{1}{B} \hat{f}_i.
\end{aligned}
\end{gather}
Using the identity
\begin{equation}
\begin{gathered}
    -\frac{1}{2B} J^\prime(r) \hat{r}^2 + \frac{\mu}{4B} \hat{z}^2 = \hat{r} \Pi_r + \hat{z} \Pi_z - B \mathcal{G}, \\
    \mathcal{G} = \frac{J^\prime(r) \Pi_\nu^2 - 2 J(r) \Pi_r \Pi_\nu + 2 \mu \Pi_r^2}{2 (J^2 (r) - 2 \mu J^\prime(r))} + \frac{1}{\mu} \Pi_z^2,
\end{gathered}
\end{equation}
where $\Pi_\nu = \Pi_\rho + 2 \Pi_z$, the action is expressed in terms of the momenta as 
\begin{widetext}
\begin{equation}
\begin{split}
    S = \int \biggl[& \Pi_\rho (\dot{\rho} - A \rho^\prime - A^\prime) + 2 \chi^\prime B \rho^\prime - 2 \chi^\pp B + \hat{r} \Pi_r + \hat{z} \Pi_z - B \mathcal{G} + \half J^\prime(r) {r^\prime}^2 + \frac{B}{2} e^{2\rho} J^\prime(r) - \frac{\mu}{4} B {z^\prime}^2 \\
    &+ \sum_{i=1}^N \left(\half \Pi_i \hat{f}_i - \frac{B}{2} {f_i^\prime}^2\right)\biggr] d^2 x + \int_{\mathcal{B}} 2 \chi^\prime B \, d\Sigma.
\end{split}
\end{equation}
\end{widetext}
Integrating by parts once more to remove the $A^\prime$ term, we identify the Hamiltonian as 
\begin{equation}
    H = \int_{\Sigma_t} (A \mathcal{C}_A + B \mathcal{C}_B) \ d\Sigma + [A\Pi_\rho - 2 \chi^\prime B]_{S_t},
\end{equation}
where 
\begin{equation}
\begin{gathered}
    \mathcal{C}_A = -\Pi_\rho^\prime + \Pi_\rho \rho^\prime + \Pi_r r^\prime + \Pi_z z^\prime + \half \sum_{i=1}^N \Pi_i f_i^\prime , \\
    \begin{aligned}
        \mathcal{C}_B =& -2 \chi^\prime \rho^\prime + 2 \chi^\pp + \mathcal{G} - \half J^\prime(r) {r^\prime}^2 - \frac{1}{2} e^{2\rho} J^\prime(r) \\
        &+ \frac{\mu}{4} {z^\prime}^2 + \half \sum_{i=1}^N {f_i^\prime}^2
    \end{aligned}
\end{gathered}
\end{equation}
are the primary constraints.

As the constraints vanish on-shell, the value of the Hamiltonian is given entirely by the boundary term, which will be taken to $r \rightarrow \infty$ along a spatial or outgoing null curve. In the gauge where $A=0$ and $B = e^{-\rho}$, i.e. the shift is 0 and the lapse is equal to 1, 
\begin{equation}
    H = [-2 \chi^\prime e^{-\rho}]_{S_t} = [- J r^\prime e^{-\rho} - \mu z^\prime e^{-\rho}]_{S_t}.
\end{equation}

To continue with the calculation some physical assumptions must be made about the solution. First, we assume that the spacetime approaches the static vacuum solution, \cref{eq:static_metric_tr}, at spatial infinity. Second, we assume that the spacetime is asymptotically flat. A necessary condition for the classical solution to be asymptotically flat is that $J(r)$ diverge at least as fast as $r$ when $r \rightarrow \infty$. A priori it is not guaranteed that the semi-classical spacetime will also be asymptotically flat, but numerical simulations have found this to be the case for many black hole models \cite{AshtekarEtAlEvaporationTwodimensional2011,BarenboimEtAlEvaporationRegular2025}, including all models considered in this paper. We assume it to be a general feature of semi-classical 2D dilaton gravity that the spacetime is asymptotically flat when the corresponding classical solution is asymptotically flat. These assumptions imply that $r^\prime = e^{2\rho}$ at the boundary, and further that $z=0, e^{2\rho} = \f$ at spatial infinity.

Finally, to normalize the energy to be zero in flat spacetime we subtract off the vacuum solution, $z = 0, M = 0$, which implies $\rho = 0, r^\prime = 1$. The mass is therefore given by 
\begin{equation}\label{eq:Hamiltonian_value}
    \mathcal{M} = [J (1 - e^{-\rho} r^\prime) - \mu z^\prime]_{\mathcal{B}}
\end{equation}
where we have dropped the $e^{-\rho}$ factor in the last term since $e^{-\rho} = 1 + O(r^{-1})$ at the boundary. Therefore the corrected mass is equal to the classical mass, which is a generalization of the Brown-York mass \cite{BrownYorkQuasilocalEnergy1993}, plus a correction term related to the Polyakov field $z$. We note that while the classical part will depend on the form of the dilaton action chosen, for example through a Weyl transformation or a further generalization as in \cite{KunstatterEtAlNew2D2016}, the correction term depends only on the Polyakov action and is independent of the initial model.

The ADM mass is given by taking the limit to spatial infinity. Here we have $z = 0$ and $r^\prime = \f$, so that
\begin{equation}
\begin{split}
    \mathcal{M}_{ADM} &= \lim_{r \rightarrow \infty} [J (1 - e^{-\rho} r^\prime) - \mu z^\prime] \\
    &\approx \lim_{r \rightarrow \infty} \left[J - J \left(1 - \frac{M}{J} \right)  \right] \\
    &= M
\end{split}
\end{equation}
and we find that $M$ is the ADM mass as expected. The only contribution is at $r \rightarrow \infty$ as boundary conditions require the mass function vanish at $r = 0$, consistent with the interpretation of 2D dilaton gravity as the dimensional reduction of a spherically symmetric spacetime.

The Bondi mass is obtained by taking the limit along a null line to $\scri^+$, 
\begin{equation}\label{eq:bondi_mass}
    \mathcal{M}_{\text{Bondi}}(u) = \lim_{v \rightarrow \infty} \left[J (1 - e^{-\rho} r^\prime) - \mu z^\prime \right].
\end{equation}

To see the physical interpretation of the correction term, we introduce the null conformal gauge ${ds^2 = -e^{2\rho} du \, dv}$. Then $z$ can be written as
\begin{equation}
    z = 2 \rho + z_u(u) + z_v(v).
\end{equation}
Since $z$ is a scalar, under a coordinate transformation
\begin{equation}
    ds^2 = -e^{2\rho} du \, dv \rightarrow -e^{2\hat{\rho}} d\hat{u} \, dv
\end{equation}
the function $z_u$ must transform as 
\begin{equation}
    z_{\hat{u}}(\hat{u}) = z_u(u) + \ln \frac{d\hat{u}}{du}.
\end{equation}
Consider a spacetime that is initially vacuum, with a black hole formed by collapse and subsequently evaporating to an asymptotically flat spacetime. This scenario is described in more detail in \cite{BarenboimEtAlEvaporationRegular2025}. Let $u, v$ denote coordinates that are Minkowski ($\rho = 0$) in the initial vacuum region, and $\hat{u}, v$ the coordinates that are inertial at future null infinity. We maintain this notation henceforth, and emphasize that the Bondi mass should be evaluated in the inertial frame at $\scri^+$. The condition that there is no radiation at past null infinity or before the black hole forms implies $z_u = z_v = 0$, and asymptotic flatness implies $\hat{\rho} \rightarrow 0$ at $\scri^+$. Therefore the correction term in the Bondi mass,
\begin{equation}\label{eq:zprime_zu}
z^\prime \rightarrow -\d_{\hat{u}} z_{\hat{u}} = -\d_{\hat{u}} \left(\ln \frac{d\hat{u}}{du}\right),
\end{equation}
is determined entirely by the coordinate transformation between the two frames, consistent with the common picture of Hawking radiation as arising from the mismatch between vacuum states defined by different observers. 

To calculate the flux, we begin by noting that just as the classical dilaton admits a generalized Brown-York mass, there is also a generalized Misner-Sharp mass \cite{KunstatterEtAlNew2D2016},
\begin{equation}
    \mathcal{M}_{MS} = \frac{J}{2} (1 - (\nabla r)^2),
\end{equation}
which is the conserved charge associated with the asymptotic timelike Killing vector $k^\mu = -\epsilon^{\mu\nu}\nabla_\nu r$ \cite{AbreuVisserKodamaTime2010, MaedaNozawaGeneralizedMisnerSharp2008},
\begin{equation}
    \nabla_\mu \mathcal{M}_{MS} = \epsilon_{\mu\nu} k_\alpha T^{\nu\alpha} = T_{\mu\nu} \nabla^\nu r - T \nabla_\mu r.
\end{equation}
The asymptotic behavior at future null infinity,
\begin{equation}
\begin{gathered}
    \d_v r = \half + O(J^{-1}), \quad \d_{\hat{u}} r = -\half + O(J^{-1}), \\
    e^{2\hat{\rho}} = 1 + O(J^{-1}),
\end{gathered}
\end{equation}
where by $O(J^{-1})$ we mean that the expansion in $v$ has the same leading order as $\frac{1}{J}$, implies
\begin{equation}
    \lim_{v \rightarrow \infty} J(1 - r^\prime e^{-\rho}) = \lim_{v \rightarrow \infty} \frac{J}{2} (1 - (\nabla r)^2), 
\end{equation}
and the Misner-Sharp mass evaluated at $\scri^+$ coincides with the classical part of the Bondi mass. Thus we may utilize this flux law for the Bondi mass. For an asymptotically flat spacetime the flux at $\scri^+$ becomes, in the inertial frame,
\begin{equation}
    \d_{\hat{u}} \mathcal{M}_{MS} = -\mu (\d_{\hat{u}} \d_{\hat{u}} z + \half \d_{\hat{u}} z \, \d_{\hat{u}} z).
\end{equation}
Because $z^\prime = -\d_{\hat{u}} z$ at $\scri^+$, we see that the Bondi mass, \cref {eq:bondi_mass}, satisfies
\begin{equation}\label{eq:bondi_mass_flux2}
    \d_{\hat{u}} \mathcal{M}_{\text{Bondi}} =  \d_{\hat{u}} (\mathcal{M}_{MS} + \mu \d_{\hat{u}} z) = -\half \mu (\d_{\hat{u}} z)^2.
\end{equation}
From \cref{eq:zprime_zu},
\begin{equation} 
    \d_{\hat{u}} z = \left(\frac{d\hat{u}}{du} \right)^{-2} \frac{d^2 \hat{u}}{du^2},
\end{equation}
and so \cref{eq:bondi_mass_flux2} reproduces the balance law given by ATV in eq. (8) of \cite{AshtekarEtAlInformationNot2008}.

\begin{figure*}[tbhp]
    \subfloat[Schwarzschild, $M=1, \mu=1$]{\includegraphics[width=0.48\linewidth]{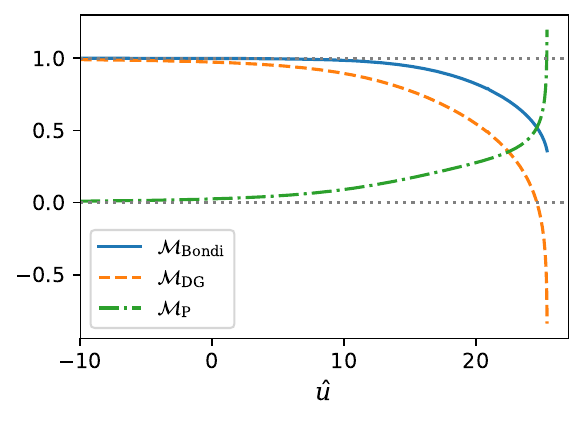}}
    \hfill
    \subfloat[Hayward, $M=0.7, \mu=2$]{\includegraphics[width=0.48\linewidth]{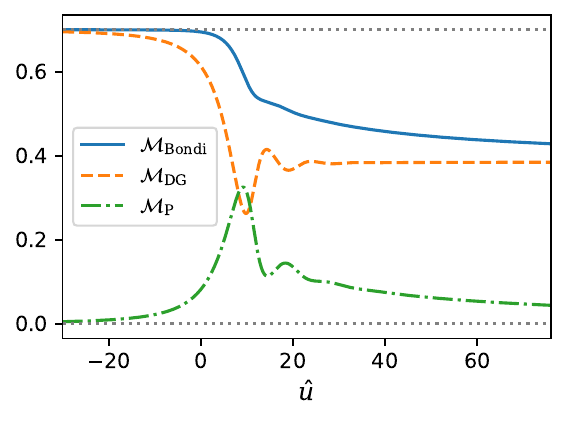}}
    \\
    \subfloat[Bardeen, $M=0.7, \mu=6$]{\includegraphics[width=0.48\linewidth]{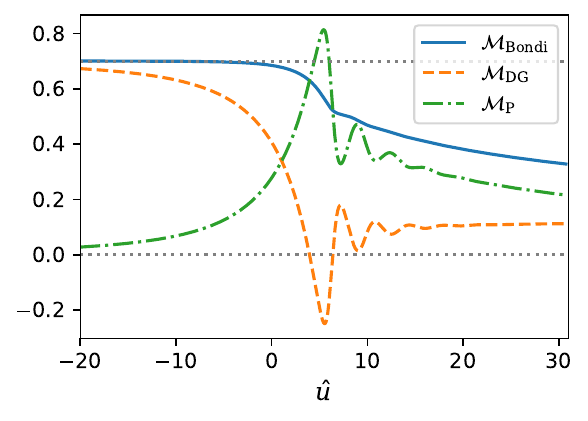}}
    \hfill
    \subfloat[Bardeen, $M=1.3, \mu=0.5$]{\includegraphics[width=0.48\linewidth]{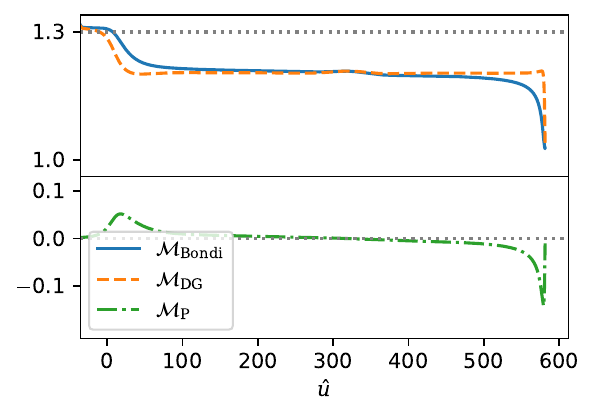}}
    \caption{The Bondi mass and its decomposition into a classical part ($\mathcal{M}_{DG}$) and quantum correction ($\mathcal{M}_P$), in asymptotically Minkowski coordinates, for different semi-classical evaporating black holes formed by the collapse of a thin null shell onto an initial vacuum. Details of the models and the numerical methods used are available in \cite{BarenboimEtAlEvaporationRegular2025}. For the Schwarzschild black hole the simulation ends at the last ray where the horizon and singularity meet. In the other models the simulation ends when the collapsing matter that generates the black hole reaches $r=0$. The continuation of the simulations past the endpoints will depend on the boundary conditions imposed there. One natural choice for the boundary conditions, proposed in the RST model \cite{RussoEtAlEndPoint1992}, is to match the evaporating solution to a new interior vacuum region. In the higher-dimensional picture this corresponds to the matter shell re-expanding outward, carrying any remaining energy back to future null infinity.}
    \label{fig:Bondi_mass_numerical}
\end{figure*}

\begin{figure*}[tbhp]
    \includegraphics[width=0.48\linewidth]{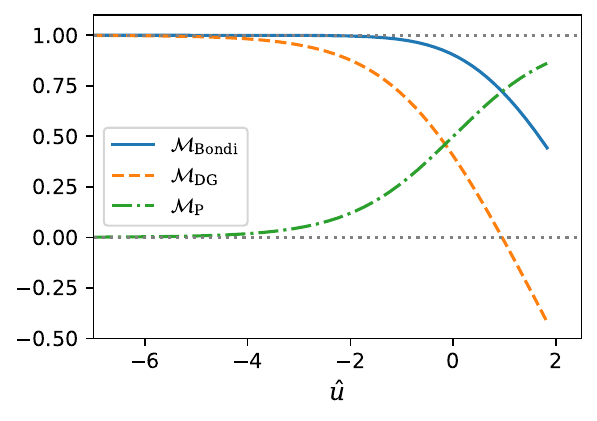}
    \includegraphics[width=0.48\linewidth]{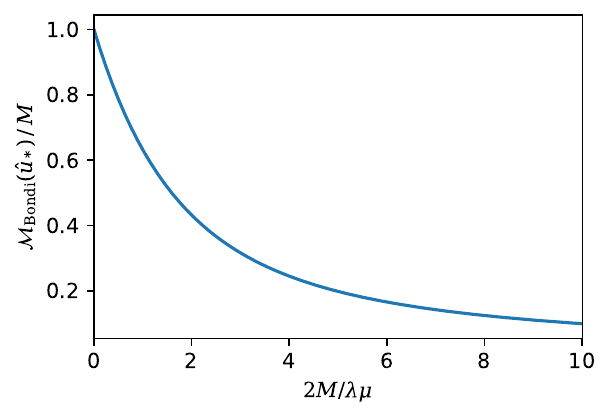}
    \caption{(Left) The Bondi mass and its decomposition for the RST model, $M = \lambda = \mu = 1$. As in the Schwarzschild model, the simulation ends at the last ray where the horizon and singularity meet. (Right) The final Bondi mass at the last ray as a function of the initial black hole mass.}
    \label{fig:Bondi_mass_RST}
\end{figure*}

From the mass flux \cref{eq:bondi_mass_flux2} we see that the Bondi mass is strictly non-increasing, and in the models we have examined we find that the Bondi mass remains non-negative. Some examples are shown in \cref{fig:Bondi_mass_numerical} and \cref{fig:Bondi_mass_RST}.

{\em In conclusion}, we have derived the Bondi mass from the Hamiltonian for generic semi-classical 2D dilaton gravity. The Bondi mass is equal to the classical mass obtained from the dilaton gravity action and a correction term obtained from the Polyakov action. The latter is determined entirely by the coordinate transformation between initial and final inertial frames, reflecting a common interpretation of Hawking radiation. The corrected Bondi mass is manifestly non-increasing in asymptotically flat spacetimes and remains positive in our numerical simulations. Positivity is demonstrated analytically for the RST model, but a general analytical proof remains for future work.

An important question is how the Bondi mass would be modified in semi-classical 4D gravity. While the classical part of the mass is closely related to the standard expressions in general relativity due to the dimensional reduction interpretation of 2D dilaton gravity, the modelling of Hawking radiation with the Polyakov action is specific to two dimensions, so the expressions derived do not generalize straightforwardly to higher dimensions. Some work has been done to study black hole evaporation in 4D \cite{LoweThorlaciusEffectiveField2025,BoyanovEtAlSemiclassicalEvolution2025a,MottolaVaulinMacroscopicEffects2006,ParentaniPiranInternalGeometry1994a,AyalPiranSphericalCollapse1997,KawaiYokokuraModelBlack2017,BalbinotEtAlAnomalyInduced1999}, but a full four-dimensional treatment is, for the moment, out of our reach.

\begin{acknowledgments}
The authors gratefully acknowledge that this research was supported in part by Discovery Grants number 2020-05346 (AF) and 2018-04090 (GK) from the Natural Sciences and Engineering Research Council of Canada.
\end{acknowledgments} 

\appendix
\section{RST model}\label{app:RST}
The RST model \cite{RussoEtAlEndPoint1992} is the only known semi-classical 2D dilaton gravity model admitting an exact analytical solution. In our formulation, the RST model is given by 
\begin{equation}
    J(r) = 2 \lambda e^{2\lambda r} + \mu \lambda
\end{equation}
with the change of variables $\phi = -\lambda r, \mu = \half \kappa$ returning the action to the more common notation for the RST action. As $\phi$ can take any real value in the RST model, we do not impose any boundary conditions at $r=0$ as in the other models, but the only contribution to the ADM and Bondi masses remains at infinity as the interior spacetime is taken to be the linear dilaton vacuum (LDV)\footnote{See \cite{deAlwisTwodimensionalQuantum1994} for a thorough discussion of the boundary conditions in RST.}. 

The solution is given by
\begin{equation}
\begin{aligned}
    r &= -\frac{1}{\lambda} \left( \frac{1}{\mu} \Omega + \half W_{-1}\left(-\frac{2}{\mu} e^{-\frac{2}{\mu} \Omega} \right) \right), \\
    \hat{\rho} &= -\lambda r + \frac{\lambda}{2} (v - \hat{u}),
\end{aligned}
\end{equation}
where 
\begin{equation}
    \Omega = e^{\lambda(v - \hat{u})} - \frac{\mu \lambda}{2} v - \frac{\mu}{2} \ln \left( e^{-\lambda \hat{u}} + \frac{M}{\lambda} \right) + \frac{M}{\lambda},
\end{equation}
and $W$ is the Lambert W or product log function. The quantum-corrected Bondi mass, in the asymptotically Minkowski coordinates, is given by
\begin{equation}
    \mathcal{M}_B (\hat{u}) = M + \frac{M \mu}{2} \frac{1}{e^{-\lambda \hat{u}} + \frac{M}{\lambda}} - \frac{\mu \lambda}{2} \ln \left( 1 + e^{\lambda \hat{u}} \frac{M}{\lambda} \right),
\end{equation}
and at the last ray,
\begin{equation}
    e^{\lambda \hat{u}_*} = \frac{\lambda}{M} \left(e^{2M/\lambda\mu} - 1 \right),
\end{equation}
the mass is
\begin{equation}
    \mathcal{M}_{\text{Bondi}} (\hat{u}_*) = \frac{\lambda \mu}{2} \left(1 - e^{-2M/\lambda \mu} \right).
\end{equation}
This is positive for positive $M, \lambda, \mu$, and because the Bondi mass is non-increasing this implies the Bondi mass remains positive throughout the evaporation process for RST black holes of any positive initial mass.

\bibliography{references}

\end{document}